\documentclass[%
 aip,
rsi,%
 amsmath,amssymb,
 reprint,%
]{revtex4-1}

\usepackage{graphicx}% Include figure files
\usepackage{dcolumn}% Align table columns on decimal point
\usepackage{bm}% bold math
\begin{document}

\preprint{AIP/123-QED}

\title[3D-Printed Chiral Metasurface]{3D-Printed Chiral Metasurface as Dichroic Dual-Band Polarization Converter}

\author{Shengzhe Wu}
\affiliation{College of Physics, Jilin University, 2699 Qianjin St., Changchun 130012, China} 
\author{Su Xu}
 \email{xusu@jlu.edu.cn}
\affiliation{State Key Laboratory of Integrated Optoelectronics, College of Electronic Science and Engineering, Jilin University, 2699 Qianjin St., Changchun 130012, China}
\affiliation{International Center of Future Science, Jilin University, 2699 Qianjin St., Changchun 130012, China}
\author{Tatiana L. Zinenko}
\affiliation{O.~Ya.~Usikov Institute for Radiophysics and Electronics of National Academy of Sciences of Ukraine, 12, Proskury St., Kharkiv 61085, Ukraine}
\author{Vladimir~V.~Yachin}
\affiliation{International Center of Future Science, Jilin University, 2699 Qianjin St., Changchun 130012, China}
\affiliation{Institute of Radio Astronomy of National Academy of Sciences of Ukraine, 4 Mystetstv St., Kharkiv 61002, Ukraine}
\author{Sergey~L.~Prosvirnin}
\affiliation{Institute of Radio Astronomy of National Academy of Sciences of Ukraine, 4 Mystetstv St., Kharkiv 61002, Ukraine}\author{Vladimir~R.~Tuz}
 \email{tvr@jlu.edu.cn}
\affiliation{State Key Laboratory of Integrated Optoelectronics, College of Electronic Science and Engineering, Jilin University, 2699 Qianjin St., Changchun 130012, China}
\affiliation{International Center of Future Science, Jilin University, 2699 Qianjin St., Changchun 130012, China}
\affiliation{Institute of Radio Astronomy of National Academy of Sciences of Ukraine, 4 Mystetstv St., Kharkiv 61002, Ukraine}

\date{\today}% It is always \today, today,
             %  but any date may be explicitly specified

\begin{abstract}
We propose a novel design of a true 3D chiral metasurface behaving as a spatial polarization converter with asymmetric transmission. The metasurface is made of a lattice of metallic sesquialteral (one and a half pitch) helical particles. Each particle contains six rectangular bars arranged in a series one above the another creating a spiral. The proposed metasurface exhibits a dual-band asymmetric transmission accompanied by the effect of a complete polarization conversion in the response on the particular distributions of currents induced in the particle's bars by an incident wave. Regarding circularly polarized waves the metasurface demonstrates a strong circular dichroism. A prototype of the metasurface is manufactured for the microwave experiment by using 3D-printing technique utilizing Cobalt-Chromium alloy, which exhibits good performances against thermal fatigue and corrosion at high temperatures. Our work paves the way to find an industrial solution on fabricating communication components with efficient polarization conversion for extreme environments.
\end{abstract}

% 41.20.Jb	Electromagnetic wave propagation; radiowave propagation 
% 42.25.Bs	Wave propagation, transmission and absorption           
% 78.67.Pt	Multilayers; superlattices; photonic structures; metamaterials 
                            

\maketitle

An optical diode (isolator\cite{jalas2013and, Mutlu_PhysRevLett_2012}) is a device transmitting light in one direction and blocking it in the reverse direction (i.e., it demonstrates an asymmetric one-way transmission feature). This characteristic appears from the effect of optical \textit{nonreciprocity} which is usually related to the time-reversal symmetry breaking of the light-matter interaction (the appearance of this effect is covered by the Lorentz lemma and reciprocity theorem\cite{landau_1960_8, potton}).

Among various types of optical diodes, one well-known example of the nonreciprocal response is the magneto-optical (Faraday) effect related to circularly polarized light propagating in gyrotropic materials.\cite{Gurevich_book_1963} The biased magnetic field breaks the time-reversal symmetry, which leads to a nonreciprocal response of media. This mechanism is accompanied by the effect of circular dichroism related to differential absorption of the left-handed and right-handed circularly polarized waves. However, the nonreciprocal response is not restricted to the use of magnetic field only. Indeed, optical diodes can be constructed by utilizing spatial-temporal modulations of refractive indices\cite{yu2009complete} or some nonlinear effects\cite{Kivshar_APL_2010, Tuz_JOSAB_2011, Lepri_PhysRevLett_2011}  for which the reciprocity theorem is violated.

Alternatively, an asymmetric transmission can be realized in chiral artificial structures (metamaterials) even when they are completely reciprocal. In such structures the effect of an asymmetric transmission arises from the spatial symmetry breaking in the particles forming the metamaterial.\cite{Collins_AdvOptMat_2017} It appears due to the specific near-field distribution and/or coupling between particular modes induced in chiral particles by an incident wave. From the viewpoint of group-theoretical description,\cite{Dmitriev_Metamat_2011, Dmitriev_IEEEAntennas_2013} if the four-fold ($C_4$) symmetry of the unit cell is broken, then the metamaterial demonstrates an asymmetric transmission even for a normally incident linearly polarized wave.\cite{Huang_PhysRevB_2012, Menzel_PhysRevA_2010, Niemi_Science_2009} This phenomenon is similar to the Faraday effect in magneto-optical materials,\cite{Floess_RepProgPhys_2018} but with no need for an external magnetic field.\cite{Singh_PhysRevB_2009} 

Typical 2D chiral planar metamaterials (metasurfaces) do not change the polarization state of the normally incident wave in the main diffraction order.\cite{Prosvirnin_PhysRevE_2005, plum2010asymmetric} Therefore, in order to obtain an asymmetric transmission in the \textit{subwavelength} structure one should construct a quasi-2D planar chiral metamaterial.\cite{Polevoy_EurPhysJApplPhys_2013} They are usually made by using 2D planar technologies arranging metallic patterns on two sides of a dielectric layer to form bilayered metasurfaces or ordering such layers into a stack in multilayered configurations. Such structures are widely used in the microwave spectral range,\cite{Han_ApplPhysLett_2011, Mutlu_PhysRevLett_2012, Shi_ApplPhysLett_2013} although similar designs are also realized for both terahertz and visible spectral ranges.\cite{Grady_Science_2013, Pfeiffer_PhysRevLett_2014} 

However, two limitations are peculiar to chiral metamaterials based on 2D planar technologies. The first limitation is related to their electromagnetic properties considering that a complete asymmetric transmission is a resonant feature strongly depended on dissipation.\cite{RDCSS0735272717050016} The second limitation is that the commercial glass epoxy laminates cannot provide a mechanical support of metallic patterns at extremely high temperatures, which restricts their application area. 

Regarding true 3D chiral metamaterials, the particles forming such structures are quite complicate objects (a true 3D chiral object is impossible to superimpose onto its mirror image even if it is lifted from its plane), so their fabrication poses a significant challenge. Fortunately, modern innovative technologies give amazing opportunities for production of true 3D chiral metamaterials,\cite{Soukoulis_ACSPhotonics_2015, Wang_APL_2018} and complex metamaterial systems with compensated chirality. \cite{doi:10.1063/1.4973679} In particular, 3D-printing techniques allow producing structures with a complex spatial topology implementing meta-devices with a new functionality.\cite{Camposeo_AdvOptMat_2018} It opens a prospects for rapid prototyping and final production of numerous diverse types of devices for the manipulation of light. Moreover, the selective laser melting techniques makes it possible to perform 3D-printing directly in solid metals to realize complex conductive devices able to operate at relatively high temperatures.\cite{Headland_OptExpress_2016} 
 
In this Letter, we propose and study a novel design of a true 3D chiral metasurface, whose prototype is fabricated by using the direct 3D printing from a metallic alloy. In the difference from the structure of work,\cite{PhysRevX.5.031005} we placed spiral particles with their axes oriented transversely to metasurface plane. The resulting structure performs a dual-band asymmetric transmission accompanied by a complete polarization conversion of the incident linearly polarized wave. The metasurface demonstrates a broadband circular dichroism of circularly polarized waves. In order to reveal the underlying mechanism of the asymmetric transmission, we perform calculation of the distribution of currents induced in particles by the incident wave. All distinctive features of the proposed metasurface are verified in the microwave experiment.

\begin{figure}[t!]
\centering
\includegraphics[width=0.95\linewidth]{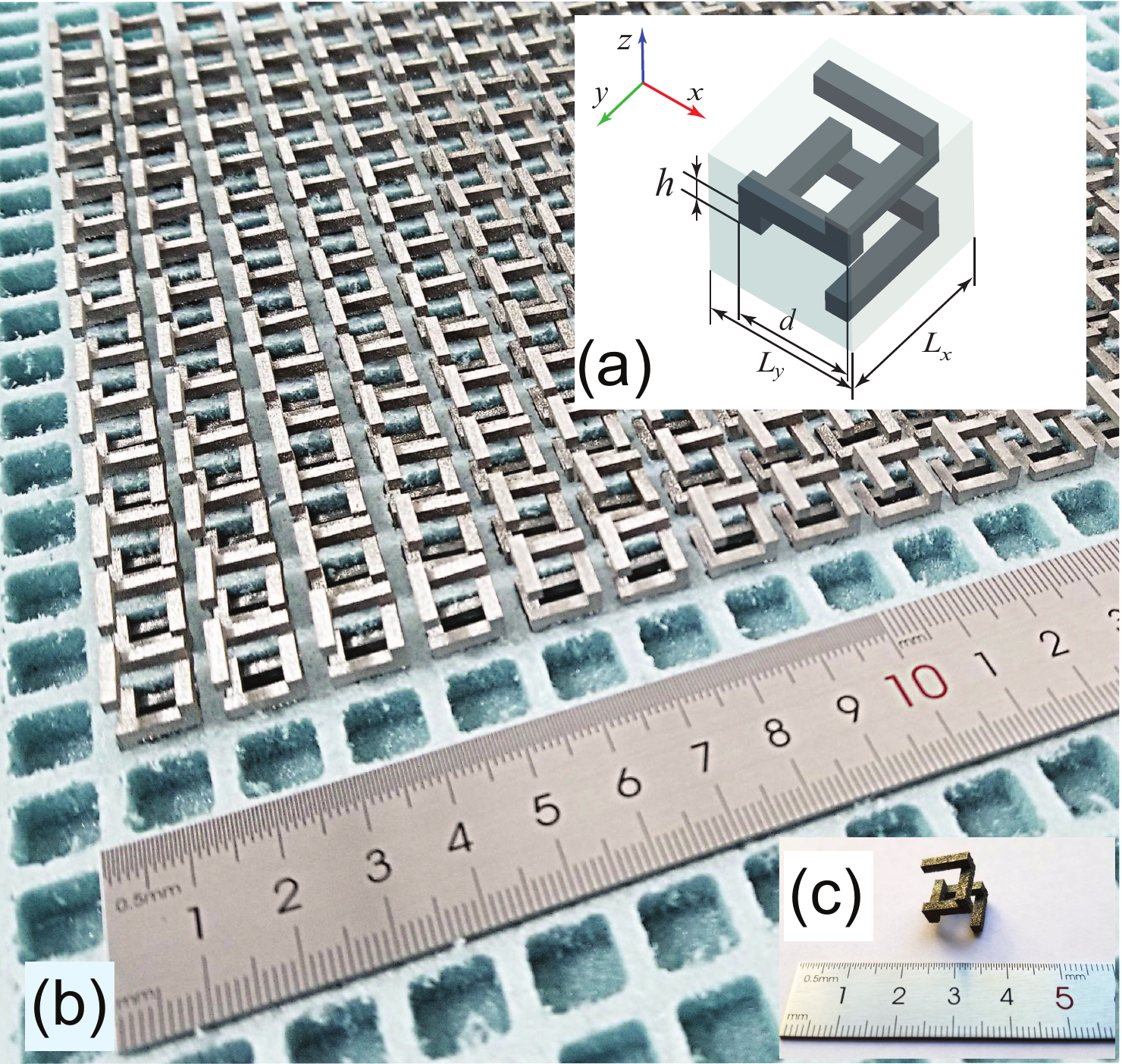}
\caption{(a) A sketch of the unit cell, (b) sample, and (c) single chiral particle of the  metasurface under study. The particle is composed of six rectangular metallic bars, where $h=2.1$~mm and $d=10.1$~mm are the bar's thickness and length, respectively.  All particles are arranged equidistantly into a lattice with period $L_x = L_y = 13.2$~mm. The lattice is fixed in a custom holder having permittivity $\varepsilon_s = 1.1$ and thickness $h_s=20.0$~mm to form a whole metasurface.}
\label{fig:sample}
\end{figure} 

The metasurface under study is made of identical metallic particles structured in the form of a spiral (helix). The particles are arranged in a periodic array and buried into a dielectric host. Each spiral particle is formed by stacking six bars one upon another. The bars are square in the cross section. The thickness of the bars is much less than their linear length. The square base of the resulting spiral particle is comparable to the lattice constant providing the subwavelength restriction on the lattice constant holds ($L/\lambda<0.5$, where $\lambda$ is the wavelength of the irradiating wave). A sketch of the periodic cell and all parameters of the metasurface are given in Fig.~\ref{fig:sample}(a). 

The metasurface was fabricated for its characterization in the microwave spectral range ($3-13$~GHz). We have manufactured a set of chiral particles from a commercially available Cobalt-Chromium alloy (melting point is 1330 $^\circ$C) by utilizing direct 3D-printing in solid metal. Printed particles appear in a very good condition having high strength. They do not require any additional mechanical or chemical post-processing treatment. To arrange the particles into a lattice, an array of rectangular holes was milled in a custom holder made of a Styrofoam material (see Figs.~\ref{fig:sample}(b) and \ref{fig:sample}(c)). The metasurface prototype was constructed of $15 \times 15$ unit cells (so we used 225 chiral particles in total). The spectral characteristics of a single sheet of the metasurface were measured in an anechoic chamber under normal incidence conditions using broadband horn antennas (see the scheme of experimental setup presented in Ref. \onlinecite{Polevoy_EurPhysJApplPhys_2013}).

In order to get deep inside into the physics of process, we have performed a numerical simulation of the scattering electromagnetic wave by the metasurface. We use the custom method based on the derivation of a set of coupled volume integral equations related to the equivalent electric and magnetic polarization currents induced in the particles by an incident wave, while the lattice periodicity is accounted by the Floquet boundary conditions.\cite{Yachin_JOSAA_2007, Yachin_UKRCON_2017} It is assumed in the method that the particles are made of dielectric with constitutive parameters satisfying the conditions $|\varepsilon| \gg 1$ and $\varepsilon\mu\simeq 1$. Under such conditions the material of particles resembles the PEC characteristics in the model. %In our computation procedure the number of harmonics in the Floquet series expansions is fixed to be 27 along each transverse direction. 

\begin{figure*}[t!]
\centering
\includegraphics[width=0.74\linewidth]{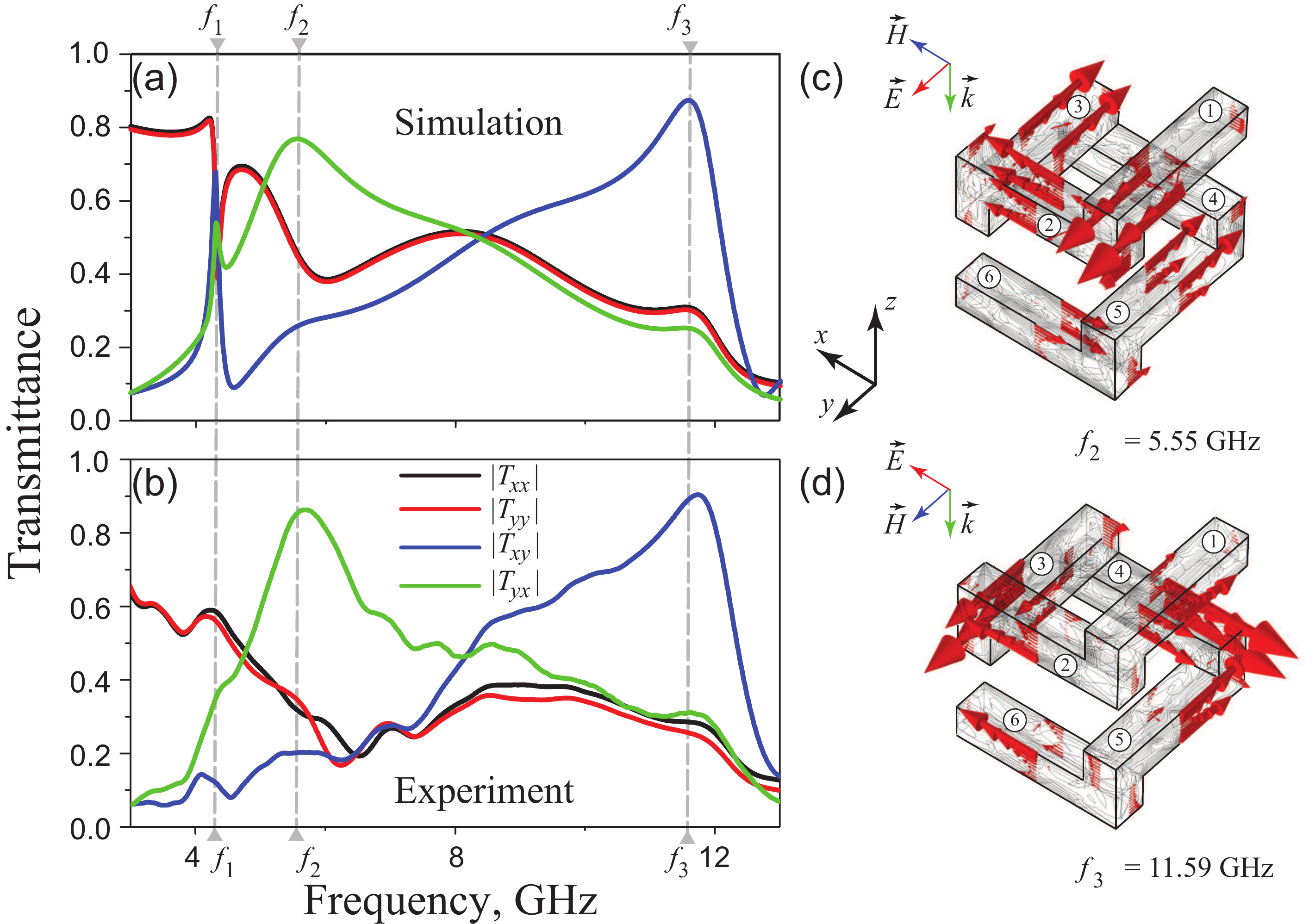}
\caption{(a) Simulated and (b) measured transmittance of the true 3D metamaterial composed of chiral particles. The structure is under an illumination of the linearly polarized wave. The directions of the polarization currents flowing along the bars forming the particle at the resonant frequencies (c)~$f_2$ and (d)~$f_3$.}
\label{fig:sim_meas}
\end{figure*}

From the obtained simulation and measurement data we have derived the transmission coefficients for both co-polarized ($T_{xx}$, $T_{yy}$) and cross-polarized ($T_{xy}$, $T_{yx}$) electromagnetic waves. These coefficients are elements of the Jones matrix $\mathbf{T}$ which relates the polarization states of the incident waves (in) to the output waves (out) in the far-field
\begin{equation}
\begin{pmatrix}
E_x^{\textrm{out}} \\
E_y^{\textrm{out}} 
\end{pmatrix} =
\mathbf{T}
\begin{pmatrix}
E_x^{\textrm{in}}\\
E_y^{\textrm{in}} 
\end{pmatrix} =
\begin{pmatrix}
T_{xx} & T_{xy} \\
 T_{yx}& T_{yy}
\end{pmatrix}
\begin{pmatrix}
E_x^{\textrm{in}}\\
E_y^{\textrm{in}} 
\end{pmatrix}. 
\label{eq:asymtrans}
\end{equation}
The frequency dependencies of the magnitude of these transmission coefficients (transmittance) are plotted in Figs.~\ref{fig:sim_meas}(a) and \ref{fig:sim_meas}(b) for both co-polarized and cross-polarized waves.

One can conclude that there is a good agreement between the results of our numerical simulation and experiment, whereas their minor discrepancy can be explained by fabrication tolerances. Another reason of deviation  in  the  results  is  that the actual  structure  has  a  finite  extent whereas the  structure  considered  in  the  simulation is expanded to  infinity. Moreover,  according  to  the  theory  predictions,  the  co-polarized  transmission coefficients have to be coincident  ($|T_{xx}|=|T_{yy}|$), whereas  in  the  experiment  there  is   slight  difference  between them due  to  the  reasons  mentioned  above.  Nevertheless, the cross-polarized transmittance of orthogonally polarized waves appears to be extremely different in the whole frequency band of interest ($|T_{xy}|\ne|T_{yx}|$) in both simulation and experiment. Further we distinguish three resonant states that appear at the particular frequencies $4.32$~GHz, $5.55$~GHz, and $11.59$~GHz. These resonant frequencies are denoted in Fig.~\ref{fig:sim_meas} by symbols $f_1$, $f_2$, and $f_3$, respectively. 

The first state at the frequency $f_1$ arises in the simulation as a strong high-$Q$ resonance, where a significant polarization conversion in the transmitted field occurs. From the analysis of distribution of currents induced on the particles by the incident wave, this resonance is a result of the out-of-phase current oscillations in the bars forming the spiral (this distribution is not presented here). Apparently, characteristics of this resonance resemble those of a trapped mode,\cite{Fedotov_PhysRevLett_2007} which is caused by the asymmetry in the particle with respect to the direction of the electric field vector of the incident wave. In the spectra of the metasurface the trapped mode typically manifests itself as a peripheral lowest frequency resonance.\cite{Tuz_OptExpress_2018} Nevertheless, this resonance completely diminishes in the transmitted spectra of the actual structure, so we exclude this resonance from our further consideration.

The remaining two states in the frequency band of interest are also recognized from the high level of polarization conversion in the transmitted field, although these resonances are much broader than the first one. At the frequencies $f_2$ and $f_3$ the cross-polarized coefficients $|T_{yx}|$ and $|T_{xy}|$ reach their maximal values, respectively, whereas the magnitudes of corresponding co-polarized coefficients $|T_{yy}|$ and $|T_{xx}|$ are small enough. Therefore, at these frequencies the given metasurface performs almost complete $y$-to-$x$ and $x$-to-$y$ conversion, when transmits the linearly polarized incident wave. This effect is similar to that found in the quasi-2D (bilayered) chiral metamaterial.\cite{Shi_ApplPhysLett_2013} 

In order to reveal the essence of these two resonant states we performed calculation of the distribution of currents induced on the particle's bars by the incident wave (for further reference we have enumerated these bars in Figs.~\ref{fig:sim_meas}(c) and \ref{fig:sim_meas}(d) by numbers \textcircled1-\textcircled6 in a series from top to bottom). It turns out that the upper part of the helix (bars \textcircled1-\textcircled3) is mainly responsible for the resonance $f_2$, while the resonance $f_3$ occurs on the lower part of the helix (bars \textcircled3-\textcircled6).

Indeed, at the resonant frequency $f_2$ there is a strong current flow along the bar \textcircled2, and this flow is directed orthogonally to the electric field vector of the incident wave (Fig.~\ref{fig:sim_meas}(c)). Such a current flow excites the cross-polarized component in the transmitted wave. Moreover, at the bars \textcircled1 and \textcircled3 also there are strong current flows, but they are directed out-of-phase to each other, which suppresses the co-polarized component of the transmitted wave. The similar situation is for the resonance at the frequency $f_3$. Here there is a strong current flow along the bar \textcircled3, and compensated current flows along bars \textcircled4 and \textcircled6, which produces the cross-polarized component and suppress the co-polarized component in the transmitted field, respectively (Fig.~\ref{fig:sim_meas}(d)).  

\begin{figure}[t!]
\centering
\includegraphics[width=0.95\linewidth]{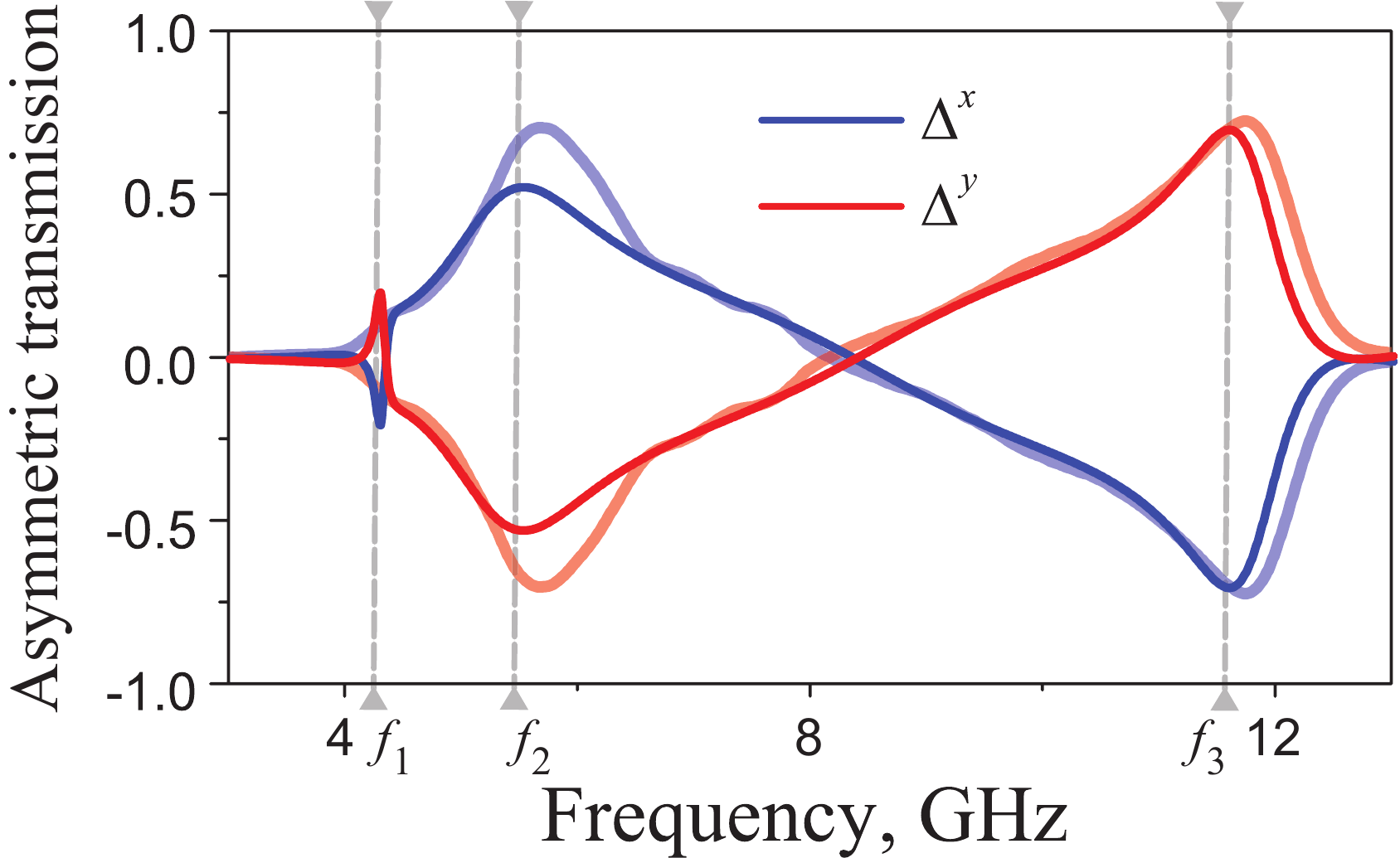}
\caption{Simulated (bright lines) and measured (pale lines) asymmetric transmission of the linearly $x$-polarized (blue lines) and $y$-polarized (red lines) incident waves.}
\label{fig:asym_trans}
\end{figure} 

\begin{figure}[t!]
\centering
\includegraphics[width=0.95\linewidth]{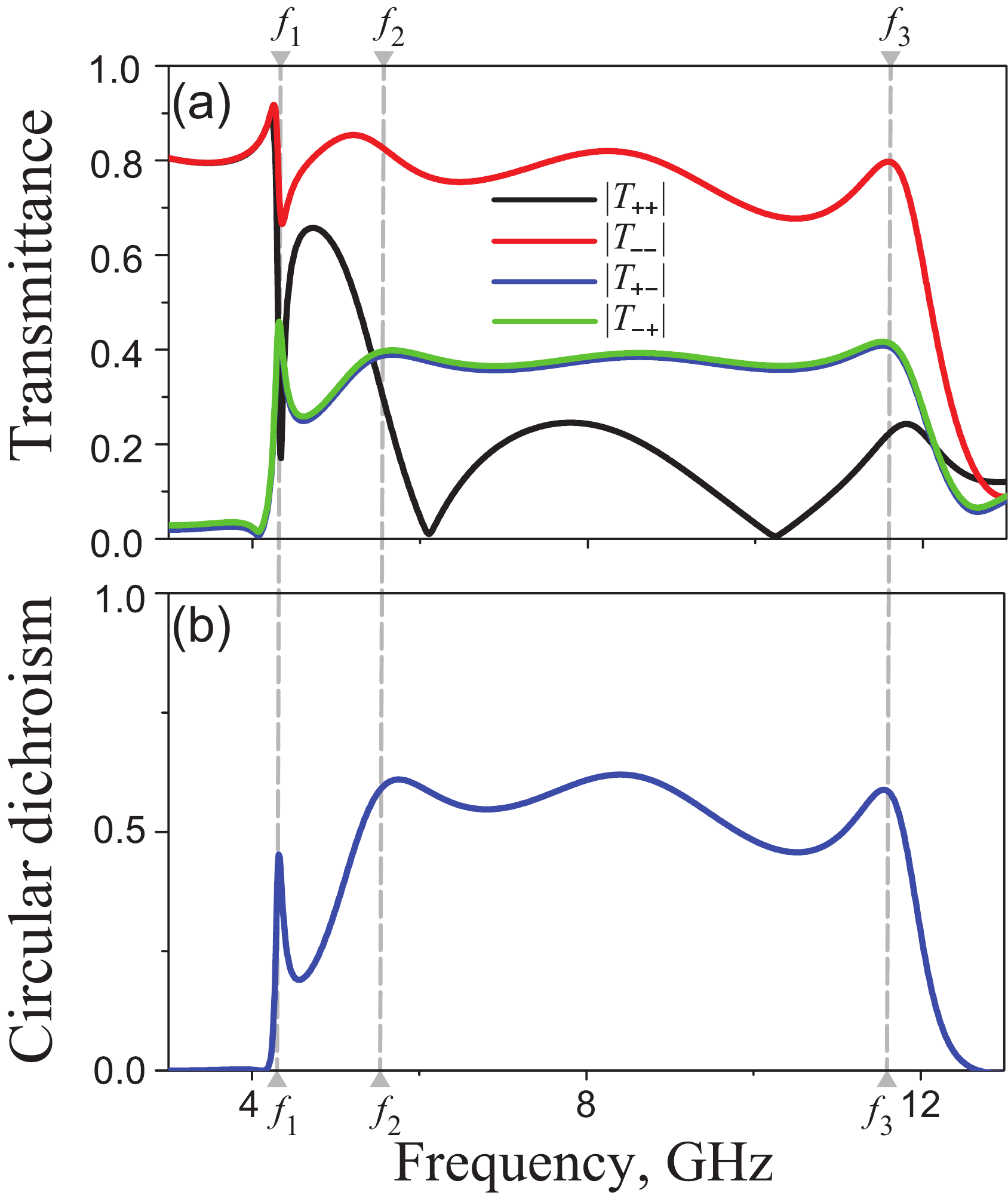}
\caption{(a) Simulated transmittance of the circularly polarized waves, and (b) broadband circular dichroism of the metasurface.}
\label{fig:sim_dc}
\end{figure} 

The asymmetric transmission parameters for the linearly $x$-polarized ($\Delta^x$) and $y$-polarized ($\Delta^y$) waves are defined as the total intensity difference for the waves propagating through a system in two opposite directions. These parameters can be alternatively expressed via two cross-polarized transmission coefficients defined for the same propagation direction as follow\cite{Menzel_PhysRevLett_2010}
\begin{equation}
\Delta^x = |T_{xy}|^2 - |T_{yx}|^2 = -\Delta^y.
\label{eq:asymtrans}   
\end{equation} 
These parameters are presented in Fig.~\ref{fig:asym_trans} as functions of frequency derived from both simulated and measured data. 

One can see that the given metasurface demonstrates a dual-band asymmetric transmission feature, where in the first band (near the frequency $f_2$) it is transmission-allowed for the $x$-polarized wave, while in the second band (near the frequency $f_3$) it is transmission-allowed for the $y$-polarized wave. The transmission of waves with the corresponding orthogonal polarization is forbidden in these bands. The experimentally measured asymmetric transmission parameters reach magnitudes of $\pm 0.74$ and $\mp 0.75$ at the frequencies $f_2$ and $f_3$, respectively (upper sign ``$+$'' is related to the $x$-polarized wave, while the lower sign ``$-$'' is related to the $y$-polarized wave).

Since the metasurface under study is a true 3D chiral material, it is able to demonstrate the effect of  circular dichroism. Remarkably, the presence of this feature distinguishes the given metasurface from the quasi-2D chiral metamaterial considered earlier.\cite{Shi_ApplPhysLett_2013}

The representation of the matrix $\mathbf{T}$ from the basis of linearly polarized waves to the basis of circularly polarized waves can be written as follow 
\begin{equation}
\begin{split}
\mathbf{T} =
\begin{pmatrix}
T_{++} & T_{+-} \\
T_{-+} & T_{--}
\end{pmatrix} &=
\frac{1}{2}
\begin{pmatrix}
T_{xx}+ T_{yy} & T_{xx} - T_{yy}\\
T_{xx} - T_{yy}& T_{xx}+ T_{yy}
\end{pmatrix}
 \\
&+\frac{i}{2}
\begin{pmatrix}
T_{xy} - T_{yx}&-T_{xy} - T_{yx}\\
T_{xy} + T_{yx}& T_{yx}-T_{xy}
\end{pmatrix}
\end{split}  
\label{eq:rcplcp}
\end{equation}
where the lower signs ``$+$'' and ``$-$'' are related to the right-handed (RCP) and left-handed (LCP) circularly polarized waves, respectively. In particular, Fig.~\ref{fig:sim_dc}(a) shows the simulated transmittance of RCP and LCP waves of the given metasurface. Since the condition $T_{+-}=T_{- +}$ holds, there is no asymmetric transmission for the circularly polarized waves. 

The differential transmittance between RCP and LCP waves defines the circular dichroism which can be expressed as \begin{equation}
\textrm{CD} = |T_{++}|^2 - |T_{--}|^2.  
\label{eq:circulardich}   
\end{equation}
Fig.~\ref{fig:sim_dc}(b) confirms the presence of a strong broadband circular dichroism in the frequency band between the resonant frequencies $f_2$ and $f_3$. Its level is about $0.6$ for the metasurface under study.

To conclude, we have proposed a novel design of a true 3D chiral metasurface, whose particles were fabricated by direct 3D printing directly in metal. The Cobalt-Chromium alloy, whose melting temperature is high, can prevent the metasurface structure from being destroyed at extreme working environment. This is a unique advantage of the alloy-based metasurface compared to the chiral components patterned on glass epoxy laminates. A dual-band asymmetric transmission feature as well as a strong broadband circular dichroism in the proposed metasurface have been demonstrated both theoretically and experimentally. We argue that a complete polarization conversion can be reached by solving a geometrical optimization problem for the required frequency band. In such a way, our work paves the way to find an industrially high-temperature component solution on polarization conversion an separating communication channels utilizing features of the linear asymmetric transmission and strong circular dichroism at microwaves as well as at higher frequencies.

VVY and VRT acknowledge Jilin University's hospitality and financial support.

%\nocite{*}
\bibliography{chiral}% Produces the bibliography via BibTeX.

\end{document}